\documentclass[pra, 12 pt]{revtex4}
\usepackage[english]{babel}
\usepackage{amsmath}
\usepackage{epsfig}

\begin{document}

%\draft

\title{\bf FROM THE DENSITY FUNCTIONAL THEORY TO THE SINGLE-PARTICLE GREEN FUNCTION
}
\author{V.B. Bobrov $^{1,2}$}
\address{$^1$ Joint\, Institute\, for\, High\, Temperatures, Russian\, Academy\,
of\, Sciences, Izhorskaia St., 13, Bd. 2. Moscow\, 125412, Russia;\\
email:\, vic5907@mail.ru\\
$^2$ National Research University "MPEI"\,,
Krasnokazarmennaya str. 14, Moscow, 111250, Russia}

\begin{abstract}
An analysis shows that the ground state of the inhomogeneous system of interacting electrons in the static external field, which satisfies the thermodynamic limit, can be consistently described only using the Green function theory based on the quantum field theory methods (perturbation theory diagram technique). In this case, the ground state energy and inhomogeneous electron density in such a system can be determined only after calculating the single-particle Green function.\\

PACS number(s): 31.15.E-, 71.15.Mb, 52.25.-b, 05.30.Fk

\end{abstract}

\maketitle

Although more than 50 years elapsed from the time of the publication of the known paper by Hohenberg-Kohn [1], the principal problem of the density functional theory (DFT) on the procedure for calculating the universal density functional has not yet been solved (see [2-4] and references therein). Therefore, in the study of the electronic structure of matter, the reduced density matrix functional theory (RDMFT) [5-7], dynamical mean-field theory (DMFT) [9,10], and Green function theory (GFT) for the many-body system [11-14] gained wide acceptance in recent years. In this situation, the problem of the relation between these theories quite naturally arises [15-17]. To solve this problem, let us consider the DFT fundamentals.

According to the Hohenberg-Kohn lemma which is often referred to as the first Hohenberg-Kohn theorem, the same inhomogeneous density  $n({\bf r})$ cannot correspond to two different local potentials $v_1({\bf r})$  and $v_2({\bf r})$   of the external field in the ground state of the non-relativistic electronic system (except for the case $ v_1({\bf r}) - v_2({\bf r}) = \mathrm{const} $)  [1]. This statement is mathematically rigorous and is beyond question.
Thus, the inhomogeneous density $n({\bf r})$   of the non-relativistic ground-state electronic system uniquely corresponds to the external field potential  $v({\bf r})$  (to within an additive constant). In other words, the inhomogeneous density $n({\bf r})$   in the ground state is the unique functional of the external field  $v({\bf r})$  (to within an additive constant),
\begin{eqnarray}
n({\bf r})\equiv\langle \Psi^+({\bf r})\Psi({\bf r})\rangle_0=n({\bf r},[v]).
 \label{F1}
\end{eqnarray}
Here  $\Psi^{+}({\bf r})$ and $\Psi({\bf r})$   are, respectively, the field creation and annihilation operators for electrons (hereafter, we omit spin indices), angle brackets with zero index mean averaging over the ground state of the electronic system in the static external field  .
The next stage of the DFT construction consists in the statement that, according to Eq. (1), the external field potential   is the inhomogeneous ground-state density functional [1] (see [18] for more details),
\begin{eqnarray}
v({\bf r})=v({\bf r}, [n])+\mathrm{const}.
 \label{F2}
\end{eqnarray}
From statement (2), the Hohenberg-Kohn theorem (or the second Hohenberg-Kohn theorem) [1] immediately follows, which states that in the ground state energy
\begin{eqnarray}
E_0=\langle \hat K+\hat U \rangle_0+\int v({\bf r})n({\bf r})d^3{\bf r} \label{F3}
\end{eqnarray}
the universal density functional
\begin{eqnarray}
F[n]=\langle \hat K+\hat U \rangle_0, \label{F4}
\end{eqnarray}
can be separated. The term "universal" means that the corresponding functional does not explicitly depend on the external field potential $v({\bf r})$. Here $\hat K$  and $\hat U$  are the kinetic energy and electron interparticle interaction energy operators, respectively.
The relation (4) is a basis for practical application of the DFT [1-4,18]. As follows from the above consideration, to construct the universal density functional $F[n]$ , it is necessary to determine the explicit form of the functional $v({\bf r},[n])$  (2). However, this problem can be solved only when considering one electron in the static external field [19], since the inhomogeneous density $n({\bf r})$   in this case has the form
\begin{eqnarray}
n({\bf r})=|\varphi_0({\bf r})|^2=\varphi_0^2({\bf r}),
 \label{F5}
\end{eqnarray}
where $\varphi_0({\bf r})$  is the wave function of the ground-state electron with energy   in the static external field which satisfies the Schr\"{o}dinger equation
\begin{eqnarray}
(\hat K+\hat U )\varphi_0({\bf r})=\varepsilon_0 \varphi_0({\bf r}),
 \label{F6}
\end{eqnarray}
and, without loss of generality, is a real function.
In this case, it is clear that relation (2) takes the form
\begin{eqnarray}
v({\bf r})=v[n({\bf r})]+\mathrm{const}.
 \label{F7}
\end{eqnarray}
When considering more than two noninteracting electrons, as shown in [19], the functional $v({\bf r}, [n])$  cannot be constructed. The existence of the universal density functional for two noninteracting electrons is caused by electron energy degeneracy with respect to the spin quantum number in  the non-relativistic consideration.
In the general case, the initial functional $n({\bf r}, [v])$  is nonlinear in the external field $v({\bf r})$ . This means that, without violation of the Hohenberg-Kohn lemma, the two possibilities are allowed.

 (i)  The inverse problem on the determination of the dependence of $v({\bf r})$ on  $n({\bf r})$ has individual solutions for each pair of functions $n({\bf r})$  and  $v({\bf r})$ (or for certain types (classes) of function pairs  $n({\bf r})$ and  $v({\bf r})$).

(ii)  The inverse problem has the universal solution  $v({\bf r})=v({\bf r},[n])$.

In general, this dilemma is not considered; however, it is assumed that the universal solution $v({\bf r},[n])$  takes place, which is valid for any external field and any number of particles [20]. The essence of the problem at hand can be expressed in other words. We introduce in the consideration the operator  $\hat P$ relating the functions $n({\bf r})$  and $v({\bf r})$
 : $n({\bf r})=\hat P v({\bf r})$. In this case, the operator $\hat P$  is such that the equality  $\hat P v({\bf r})=\hat P \{v({\bf r})+\mathrm{const}\}$ is valid. It follows from definition (1) for the inhomogeneous density $n({\bf r})$   that the operator $\hat P$  is nonlinear. Thus, the problem of determining the inverse operator $\hat P^{-1}$  relating the functions  $v({\bf r})$ and  $n({\bf r})$ as $v({\bf r})=\hat P^{-1} n({\bf r})$  has not a unique solution in the general case.
Thus, statement (2) about the existence of the density functional for the external field potential is not valid when considering the multielectron system. Therefore, the proof of the existence of the universal density functional  $F[n]$ (4) is absent although the Hohenberg-Kohn lemma about the functional  $n({\bf r}, [v])$  uniqueness [19,20] is valid.
Nevertheless, we would like to retain the main idea in describing the multielectron system, i.e., to use functions of a small number of spatial variables, rather than many-particle wave functions.
We take into account that the inhomogeneous density  $n({\bf r})$  is a diagonal element of the reduced density matrix
\begin{eqnarray}
\gamma ({\bf r},{\bf r'})=\langle \Psi^+({\bf r})\Psi({\bf r'})\rangle_0=\gamma ({\bf r},{\bf r'};[v]), \qquad  n({\bf r})=\gamma ({\bf r},{\bf r}).
 \label{F8}
\end{eqnarray}
Thus, the statement similar to the Hohenberg-Kohn lemma (8) for the inhomogeneous density $n({\bf r})$  is valid for the function $\gamma ({\bf r},{\bf r'})$  , i.e., the functional  $\gamma ({\bf r},{\bf r'};[v])$ is unique. Therefore, instead of statement (2), we can assume that the external field potential is the functional of the reduced density matrix,
\begin{eqnarray}
v({\bf r})=v({\bf r}, [\gamma])+\mathrm{const}.
 \label{F9}
\end{eqnarray}
In this case, instead of the universal density functional $F[n]$  (4), we obtain the universal functional of the reduced density matrix,
\begin{eqnarray}
\Phi[\gamma]=\langle \hat K+\hat U \rangle_0, \label{F10}
\end{eqnarray}

The statement (10) is a basis of the RDMFT which, in contrast to the DFT, is valid for both the arbitrary number of noninteracting electrons and for the self-consistent Hartree-Fock approximation (see [5-7] for more details).

Let us pay attention that the DFT results do not follow from the RDMFT, although the assumption that the reduced density matrix is the inhomogeneous density functional $\gamma ({\bf r},{\bf r'})=\gamma ({\bf r},{\bf r'};[n])$  is often used [5-7]. In addition to the absence of the corresponding proof, the results of the self-consistent Hartree-Fock approximation cannot be used within the DFT [21].
However, the possible existence of the universal functional of the reduced density matrix is based on statement (9) which, as in the case of (3) for the DFT, cannot be proved. As a result, we face with the absence of a regular procedure for determining the universal functional $\Phi[\gamma]$  (10) when considering the inhomogeneous system of interacting electrons.
In this context, we note that if we take the existence of the universal density functional  $F[n]$ (4) as a postulate, the existence of the functional  $v({\bf r}, [n])$ will be a strict result of the DFT [22]. In fact, this means that the existence of the density functional for the external field potential (2) follows from the existence of the universal density functional (4). It is clear that a similar statement is valid within the RDMFT as well, i.e., Eq. (9) follows from (10).
Thus, the used method for proving the existence of the universal reduced density functional $\Phi[\gamma]$  is strictly speaking incorrect. In this situation, we cannot but consider statement (10) as a postulate, at least, until proved otherwise, e.g., with respect to the universal functional  $F[n]$.
To get rid of the need to use the statements for the external field potential, similar to (2) and (9), it is necessary to specify such a function for describing the inhomogeneous electronic system in the static external field, which uniquely defines the ground state energy of the system under consideration.

The solution of this problem is possible when considering the inhomogeneous multielectron system in the static external field which satisfies the thermodynamic limit $V\rightarrow\infty$, $\langle \hat N \rangle_0\rightarrow\infty$, ${n}=\langle \hat N \rangle_0/V$, where $V=\int d^3 r$  is the volume occupied by the system,   $\langle \hat N \rangle_0=\int d^3 r n({\bf r})$ is the total number of electrons, and ${n}=\langle \hat N \rangle_0/V$  is the average number of ground-state electrons per unit volume. For such a ground-state system, based on the Gibbs grand canonical distribution with specified chemical potential $\mu$, we can introduce the time single-particle Green function
\begin{eqnarray}
g ({\bf r},t;{\bf r'},t')=\langle \hat T \{\tilde \Psi({\bf r,t}) \tilde \Psi^+({\bf r',t'})\}\rangle_0,
 \label{F11}
\end{eqnarray}
where $\tilde \Psi^+({\bf r',t'})$  and  $\tilde \Psi({\bf r,t})$  are the field creation and annihilation operators, respectively, for electrons in the Heisenberg representation with exact system Hamiltonian, and  $\hat T$ is the temporal ordering operator [23]. Hereafter, Planck's constant is $\hbar=1$ . From definition (11), it immediately follows that
\begin{eqnarray}
\gamma ({\bf r},{\bf r'})=-i \lim_{t'\rightarrow t+0}g ({\bf r},t;{\bf r'},t').
 \label{F12}
\end{eqnarray}
Hence, for the Green function $g ({\bf r},t;{\bf r'},t')$, the statement similar to the Hohenberg-Kohn lemma (8) for the inhomogeneous density $n({\bf r})$   is valid, i.e., the functional $g ({\bf r},t;{\bf r'},t';[v])$  is unique. In this case, this function is calculated using the well developed methods of the quantum field theory (the perturbation theory diagram technique for the interparticle interaction) [23]. In particular, for the Green function $g ({\bf r},{\bf r'};\omega)$  which is the Fourier transform of the function  $g ({\bf r},t;{\bf r'},t')$ with respect to the variable $(t-t')$ , the equation
\begin{eqnarray}
g^{-1} ({\bf r},{\bf r'};\omega)=g_0^{-1} ({\bf r},{\bf r'};\omega)-\Sigma ({\bf r},{\bf r'};\omega)
 \label{F13}
\end{eqnarray}
is valid, where the familiar electron self-energy $\Sigma ({\bf r},{\bf r'};\omega)$  is the functional of the exact Green function $g ({\bf r},{\bf r'};\omega)$,
\begin{eqnarray}
\Sigma ({\bf r},{\bf r'};\omega)=\Sigma ({\bf r},{\bf r'};\omega;[g]),
 \label{F14}
\end{eqnarray}
which is an infinite functional power series in the electron-electron interaction potential and single-particle Green functions, $g_0 ({\bf r},{\bf r'};\omega)$  is the Green function for the system of noninteracting ground-state electrons [23],
\begin{eqnarray}
g_0({\bf r},{\bf r'};\omega)=\sum_{\varepsilon_{0,k}>\mu}\frac{\varphi^*_{0k}({\bf r'})\varphi_{0k}({\bf r})}{\omega-\varepsilon_{0k}+i0}+\sum_{\varepsilon_{0,k}<\mu}\frac{\varphi^*_{0k}({\bf r'})\varphi_{0k}({\bf r})}{\omega-\varepsilon_{0k}-i0}.
 \label{F15}
\end{eqnarray}
Here $\varphi_{0k}({\bf r})$ and $\varepsilon_{0k}$  are the electron wave function and energy, respectively, which are defined by the single-particle Schr\"{o}dinger equation (see (6)). In this case, relation (12) takes the form
\begin{eqnarray}
\gamma ({\bf r},{\bf r'})=-i \lim_{t\rightarrow +0}\int \frac {d \omega}{2\pi}g ({\bf r},{\bf r'};\omega)\exp (i\omega t).
 \label{F16}
\end{eqnarray}
In this case, the chemical potential  $\mu$ can be found by the specified average density (taking into account the electron spin),
\begin{eqnarray}
n=n(\mu)=-\frac{2 i}{V}\lim_{t\rightarrow +0}\int d^3 r \int \frac {d \omega}{2\pi}g ({\bf r},{\bf r'};\omega)\exp (i\omega t).
 \label{F17}
\end{eqnarray}
On this basis, it can be shown [24] (see also [9]) that the quantity $\langle \hat K+\hat U \rangle_0$  is the universal functional $G([g])$  of the single-particle Green function $g ({\bf r},{\bf r'};\omega)$,
\begin{eqnarray}
G([g])=Tr (\hat K g)+\frac{1}{2}Tr ( \Sigma g).
 \label{F18}
\end{eqnarray}
Here the symbol $Tr$   denotes summation over spin indices and integration over spatial variables and frequency $\omega$  with factor $- i \exp(i \omega t)$  under the condition $t\rightarrow +0$. Using Eq. (18), we can construct the Luttinger-Word [25] and Kadanoff-Baym [26] dynamic variation procedure for determining the single-particle Green function (see [27] for more details).
Thus, using the GFT based on the quantum field theory methods, we obtain the consistent description of the ground state of the inhomogeneous electronic system in the static external field, which satisfies the thermodynamic limit, without the consideration of the problem of the functional for the external field potential (see Eqs. (2) and (9)).

In this case, an important remark should be made. As follows from Eq. (18), the functional  $G([g])$  essentially depends on the functional $\Sigma ({\bf r},{\bf r'};\omega;[g])$. At the same time, according to Eq. (13), this functional directly defines the single-particle Green function, i.e., relation (13) is a functional equation for calculating the Green functions $g({\bf r},{\bf r'};\omega)$. If we have the solution to Eq. (13), inhomogeneous density $n ({\bf r})$  is determined from Eqs. (8) and (16). In turn, to calculate the ground state energy  $E_0$, the single-particle Green function is also sufficient (see, e.g., [23]),
\begin{eqnarray}
E_0=Tr (\hat K g)+\frac{1}{2}Tr \left((\omega-\hat K+\mu)(g-g_0)\right) +\int d^3 r n ({\bf r})v ({\bf r}).
 \label{F19}
\end{eqnarray}
This means that when constructing approximations for determining the single-particle Green function, the self-consistency procedure should be performed between solutions corresponding to functional equation (13) and the Luttinger--Ward--Kadanoff--Baym dynamic variational method. Various approximations for calculating the single-particle Green function, including the relation with the DMFT, are presented in [9-17].
Thus, the ground state of the inhomogeneous electronic system in the static external field, which corresponds to the thermodynamic limit is completely defined by the single-particle Green function. Only after its calculation, the inhomogeneous density and ground state energy of the system under consideration can be determined.

\section*{Acknowledgment}

This study was supported by the Russian Foundation for Basic Research (project no. 12-08-00822-a) and the Presidium of the Russian Academy of Sciences (grant no. 2P "Substance at high energy density"). The author is grateful to S.A. Trigger for the useful discussion and remark.

\end{document}